\documentclass[prl,twocolumn,superscriptaddress,showpacs]{revtex4}

\usepackage{amsmath}
\usepackage{amssymb}
\usepackage{graphicx}
\usepackage{subfigure}

\newcommand{\be}[1]{\begin{equation}\label{#1}}
\newcommand{\ee}{\end{equation}}
\newcommand{\ba}[1]{\begin{eqnarray}\label{#1}}
\newcommand{\ea}{\end{eqnarray}}
\newcommand{\rf}[1]{(\ref{#1})}
\newcommand{\nn}{\nonumber}

\begin{document}

\title{Paradoxes of magnetorotational
instability and their geometrical resolution}




\author{Oleg N. Kirillov}
\email{o.kirillov@hzdr.de}
\affiliation{Helmholtz-Zentrum Dresden-Rossendorf\\
P.O. Box 510119, D-01314 Dresden, Germany}

\author{Frank Stefani}
\email{f.stefani@hzdr.de}
\affiliation{Helmholtz-Zentrum Dresden-Rossendorf\\
P.O. Box 510119, D-01314 Dresden, Germany}

\date{\today}

\begin{abstract}

The magnetorotational instability (MRI) triggers turbulence
and enables outward transport of angular momentum in
hydrodynamically stable accretion discs.
By using the WKB approximation and
methods of singular function theory, we resolve two different
paradoxes of MRI that appear in the limits of infinite and
vanishing magnetic Prandtl number. For the latter case
we derive a new strict limit of the critical Rossby number.
This new limit of ${\rm Ro_c}=-0.802$, which
appears for a finite Lundquist
number of ${\rm Lu}=0.618$, extends the formerly known
inductionless Liu limit of ${\rm Ro_c}=-0.828$ valid at ${\rm Lu}=0$.

\end{abstract}


\maketitle

The magnetorotational instability (MRI) is the main candidate
to explain the fast formation of stars and black holes by
triggering turbulence and angular momentum transport
in accretion disks. In its standard version (SMRI), with
a vertical field being applied, the instability is
non-oscillatory \cite{v59,c60,bh91}, while a helical
applied magnetic field
leads to an oscillatory instability (HMRI)  \cite{k92andhr05}.

Both in the astrophysical context \cite{bh98}, as well as
in laboratory experiments
\cite{lathrop04andstefani06andstefani09andnjs10}, it is vital to
know which
laws of differential rotation are susceptible to MRI.
The hydrodynamic reference point is Rayleigh's
criterion \cite{r17} stating that a rotating flow with
an outwardly increasing angular momentum is stable. This implies,
e.g., that a Taylor-Couette flow of an inviscid fluid
between the inner and outer co-axial cylinders of radii $R_{\rm i}<R_{\rm o}$
and of infinite length that rotate with
different angular velocities, $\Omega(R_{\rm i})$ and $\Omega(R_{\rm o})$, is stable,
if and only if $ R_{\rm i}^2\Omega(R_{\rm i})<R_{\rm o}^2\Omega(R_{\rm o})$.
In contrast to this, assuming a perfectly conducting fluid
and a vertical magnetic field $B_z^0$
being applied, Velikhov \cite{v59} and Chandrasekhar
\cite{c60} found the  more restrictive condition for stability
in the form $\Omega(R_{\rm i})<\Omega(R_{\rm o})$.
Remarkably, the latter criterion does not depend on the
magnetic field strength, i.e. in the limit $B_z^0\rightarrow0$ it
does not reduce to Rayleigh's criterion valid for $B_z^0=0$.
This `curious behavior of ostensibly changing the Rayleigh criterion discontinuously' \cite{B03}
constitutes the
\textit{Velikhov-Chandrasekhar paradox}
which implies a dependence of the
instability threshold on the sequence of taking the two
limits of vanishing magnetic field and
vanishing electrical resistivity.
Its physical origin has been attributed to the
fact that in a fluid of zero resistivity the
magnetic field lines are permanently
attached to the fluid, independent of the strength of the magnetic
field \cite{v59,c60}.

Another paradox of MRI emerges in the opposite
limit of vanishing
electrical conductivity. This {\textit{paradox of
inductionless HMRI}} \cite{pgg07andpgg07a} refers to
the fact that in a helical magnetic field
a perturbation can grow exponentially although
the instantaneous growth of the energy of any
perturbation must be smaller than in the field-free case.

Actually, the astrophysical relevance of HMRI is
still under debate.
On the first glance, according to the criterion
of Liu et al. \cite{lghj06},
it could only work for
rather steep rotation profiles $\Omega(R)$ with
Rossby numbers
${\rm Ro}:=R(2\Omega)^{-1} {d\Omega}/{dR}<2-2\sqrt{2} \simeq -0.828$.
This would clearly exclude any relevance of HMRI for
Keplerian profiles characterized by ${\rm Ro}=-0.75$.
It has to be noted, though, that
the validity of the underlying local WKB
approximation, and the possible role
of electrical boundaries for extending the
applicability of HMRI to higher Rossby numbers are
controversially discussed \cite{rh07}.
Surprising new arguments arose recently
from investigations of the saturation regime of MRI.
For the case of small magnetic Prandtl numbers
(as they are typical for protoplanetary disks),
Umurhan speculated about a saturated
rotation profile with regions of reduced shear, sandwiched by
regions of strengthened shear \cite{u10}. For those latter
regions with steeper than Keplerian profiles,
HMRI could indeed become of relevance.

In this Letter we  both   
find the ultimate upper limit of the critical
Rossby number for HMRI, and resolve the mentioned paradoxes.
We establish that these physical effects
sharply correspond to the geometric
singularities that are inherent
on the stability boundaries of leading-order WKB equations.

We start with the local WKB equations for the
axisymmetric perturbation of a
steady-state rotational flow of a viscous and resistive
fluid in the presence of an axial magnetic field that were
derived and discussed  by several authors
\cite{bh91,jgk01andlv07andrs08,ks10}. They can be rewritten
in the typical
form of a non-conservative gyroscopic system \cite{kv10},
\be{e1}
\ddot
u +
(D +\Omega_0(1+\alpha^2){J})
\dot
u +
(N +K)
u=0,
\ee
where  $u=(u_R,u_{\phi})^T$
is the fluid velocity in polar coordinates  $(R,\phi)$.
Separating the time-dependence according to $u=\tilde u \exp(\gamma t)$
yields the eigenvalue
problem ${L}(\gamma){\tilde u}=0$
for the growth rate of the
perturbation $\gamma$, where
${L}(\gamma)=\gamma^2{I} +\gamma({D} +\Omega_0(1+\alpha^2){J}) +{N} +{K}$,
$I$ the $2 \times 2$ unit matrix, ${N}=\Omega_0(\omega_{\eta}(1+\alpha^2)+{\rm Ro}(\omega_{\eta}-\omega_{\nu})){J}$,
\be{e2a}
{K}= \left(
            \begin{array}{ll}
              \omega_A^2+\omega_{\nu}\omega_{\eta} & k_{12} \\
              k_{12} & \omega_A^2+\omega_{\nu}\omega_{\eta}+4\alpha^2\Omega_0^2{\rm Ro} \\
            \end{array}
          \right)
\ee
with $k_{12}=\Omega_0(\omega_{\eta}(1-\alpha^2)+{\rm Ro}(\omega_{\eta}-\omega_{\nu}))$, and
\be{e2}
{J}=\left(
                                                                                                             \begin{array}{rr}
                                                                                                               0 & -1 \\
                                                                                                              1 & 0 \\
                                                                                                             \end{array}
                                                                                                           \right),~~
{ D}=\left(
          \begin{array}{cc}
            \omega_{\nu}+\omega_{\eta} & \Omega_0(1-\alpha^2) \\
            \Omega_0(1-\alpha^2) & \omega_{\nu}+\omega_{\eta} \\
          \end{array}
        \right).
\ee

In the above equations, $\omega_{\nu}=\nu k^2$ and $\omega_{\eta}=\eta k^2$
are the viscous and resistive frequencies, $\omega_A={k_z B_z^0}(\mu_0 \rho)^{-1/2}$ the Alfv\'en frequency,
$k_R$ and $k_z$ the radial and axial wave numbers of the perturbation,
$k^2=k_z^2+k_R^2$, $\alpha=k_z/k$, $\Omega_0=\Omega(R_0)$ and ${\rm Ro}={\rm Ro}(R_0)$,
where $R_0$ is the radial coordinate of a
fiducial point for
the local stability analysis. We use the convention that
$\rho=const$ is the density of the fluid, $\nu=const$ the kinematic
viscosity, $\eta=(\mu_0 \sigma)^{-1}$ the magnetic diffusivity, $\sigma$ the conductivity of the fluid,
and $\mu_0$ the magnetic permeability of free space.
For $\alpha=1$, $\omega_{\nu}=0$, and $\omega_{\eta}=0$, Eq. \rf{e1} is similar to
the Hill equation for two orbiting mass points connected by a spring
\cite{sc10},  a paradigmatic model of SMRI \cite{bh98,B03}.

Stable perturbations  have $\Re \, \gamma \le 0$ provided that
$\gamma$ with $\Re \, \gamma = 0$ is a semi-simple eigenvalue of  ${L}(\gamma)$. The growing solutions of SMRI are non-oscillatory with $\Im \gamma=0$. Therefore, $\gamma=0$ implies that $\det({N}+{K})=0$
at the threshold of SMRI \cite{bh91} which results in
the critical Rossby number (above which the flow is stable)
\ba{e4}
{\rm Ro}_{\rm c}&=&-\frac{(\omega_A^2+
\omega_{\nu}\omega_{\eta})^2+4\Omega_0^2\omega_{\eta}^2\alpha^2}
{4\Omega_0^2\alpha^2(\omega_A^2+\omega_{\eta}^2)}\nn\\
&=&-\frac{({\rm Pm}^{-1}+{\rm Lu}^2{\rm Pm}^{-2})^2+4{\rm Re}^2{\rm Pm}^{-2}}{4{\rm Re}^2({\rm Lu}^2{\rm Pm}^{-2}+{\rm Pm}^{-2})},
\ea
where
$
{\rm Re}=\alpha \Omega_0\omega_{\nu}^{-1}
$ is the Reynolds number, ${\rm Pm}=\omega_{\nu}\omega_{\eta}^{-1}$ the magnetic Prandtl number, and ${\rm Lu}=\omega_A\omega_{\eta}^{-1}$ the Lundquist number. Formula \rf{e4} coincides with that derived in \cite{ks10}
from the Routh-Hurwitz criterion \cite{Bi44}.

The Velikhov-Chandrasekhar paradox occurs at infinite ${\rm Pm}$ and means that in
the ideal MHD case ($\omega_{\eta}=0$, $\omega_{\nu}=0$) the limit $\omega_{A}\rightarrow0$
yields Velikhov's value ${\rm Ro}_{\rm c}=0$ as the instability threshold
rather than Rayleigh's limit ${\rm Ro}_{\rm c}=-1$ of the non-magnetic case $(\omega_{A}=0$, $\omega_{\nu}=0)$.

                \begin{figure}[tp]
    \centering
    \subfigure{\includegraphics[angle=0, width=0.32\textwidth]{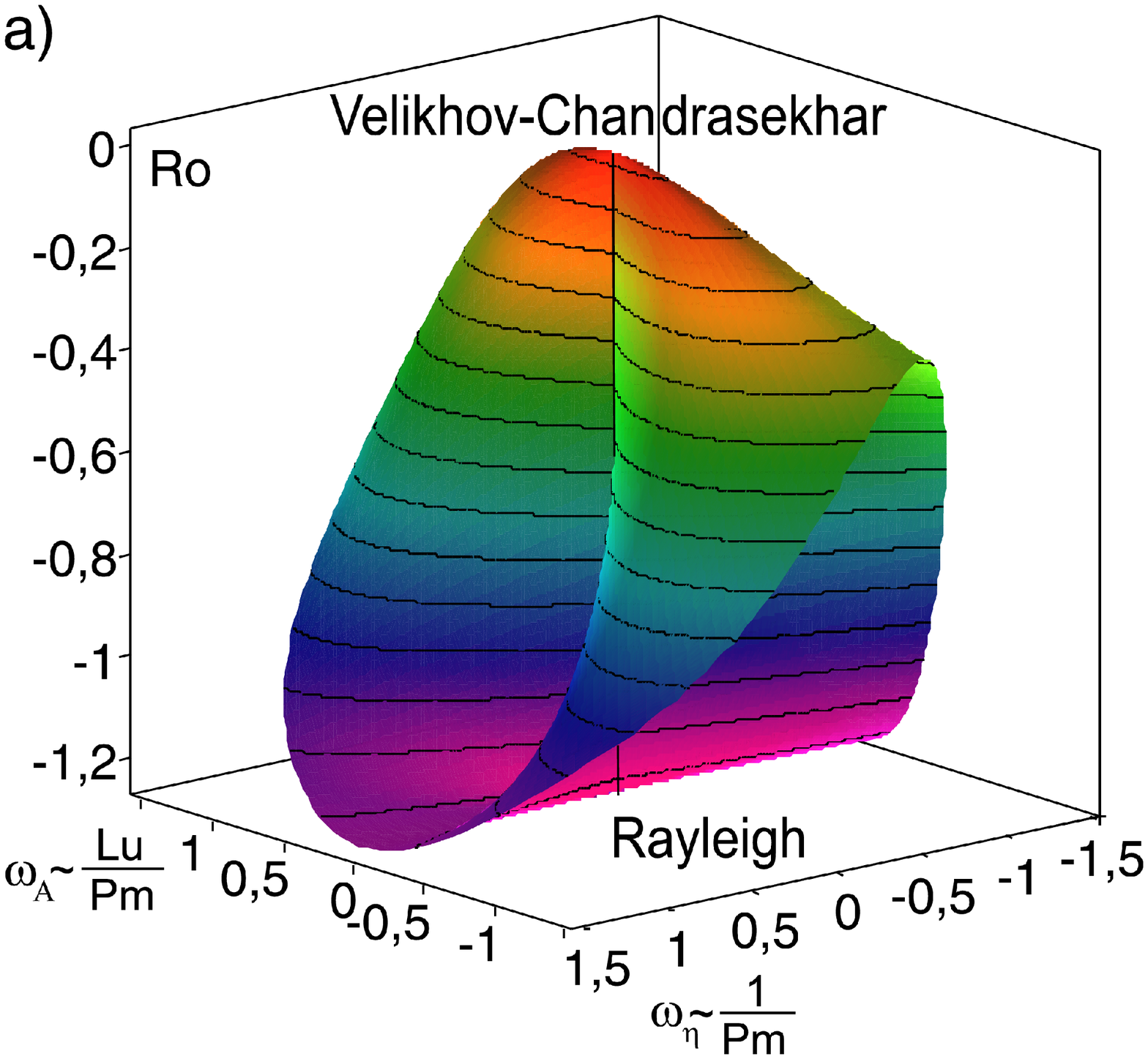}}
    \hspace{.03in}
    \subfigure{\includegraphics[angle=0, width=0.14\textwidth]{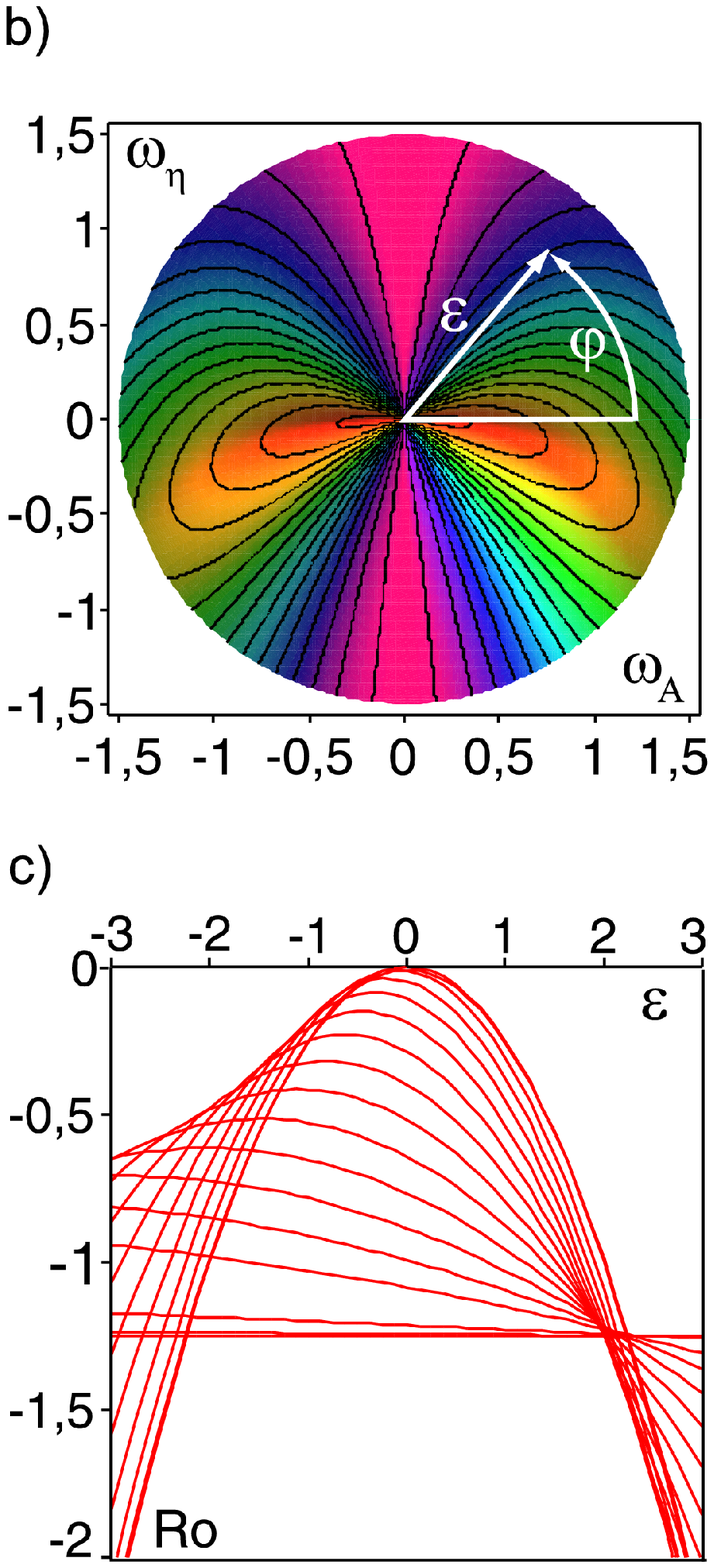}}
    \caption{(a) The critical Rossby number of SMRI as a function of  $\omega_A \sim {\rm Lu}{\rm Pm}^{-1}$ and $\omega_{\eta} \sim {\rm Pm}^{-1}$ for $\omega_{\nu}=1$, $\alpha=1$, $\Omega_0=1$, i.e. for  ${\rm Re}=1$. (b)
 Top view of the surface. (c) Cross-sections of the surface along the rays specified by the Lundquist number, or, equivalently, by the angle $\varphi$ that varies from $0$ to $1.5$ through the equal intervals $\Delta \varphi =0.1$; the horizontal line corresponds to $\varphi=\pi/2$. Note that negative values of $\omega_{\eta}$ and $\varepsilon$ are not physical.}
    \label{fig1}
    \end{figure}

With  $\omega_A=\varepsilon\cos\varphi$ and $\omega_{\eta}=\varepsilon\sin\varphi$ in \rf{e4}, we obtain
\be{e5}
{\rm Ro}_{\rm c}= -\frac{(\varepsilon\cos^2\varphi+\omega_{\nu}\sin\varphi)^2+4\alpha^2\Omega_0^2\sin^2\varphi}{4\alpha^2\Omega_0^2},
\ee
which for $\varepsilon \rightarrow 0$ reduces to
\be{e6}
{\rm Ro}_{\rm c}= -\left(1+\frac{1}{4{{\rm Re}}^2}\right)\sin^2\varphi=-\frac{1+(2{\rm Re})^{-2}}{1+{{\rm Lu}}^2}.
\ee
Introducing the new parameter ${\rm Ro}'=(1+4{{\rm Re}}^2(1+2{\rm Ro}))(1+4{{\rm Re}}^2)^{-1}$ we find that in the
$(\omega_A,\omega_{\eta},{\rm Ro}')$-space Eq. \rf{e6} defines a so-called
{\it ruled surface} $(\varepsilon,\varphi)\mapsto (\varepsilon\cos\varphi,\varepsilon\sin\varphi,\cos n\varphi)$ with $n=2$, which is a canonical equation for the Pl\"ucker conoid of degree $n=2$ \cite{hk10}. The surface
according to Eq. \rf{e4} tends to the Pl\"ucker conoid when $\varepsilon=\sqrt{\omega_A^2+\omega_{\eta}^2}$ goes to zero.
This surface is shown in the $(\omega_A,\omega_{\eta},{\rm Ro})$-space in Fig.~\ref{fig1}(a) and in projection to the $(\omega_A,\omega_{\eta})$-plane in Fig.~\ref{fig1}(b) for ${\rm Re}=1$. For each $\alpha$, $\omega_{\nu}$, and $\Omega_0$ it has the same Pl\"ucker conoid singularity, i.e. an
interval of self-intersection along the ${\rm Ro}$-axis and two Whitney umbrella
singular points at its ends. This singular structure implies non-uniqueness for the
critical Rossby number when simultaneously $\omega_A=0$ and $\omega_{\eta}=0$.
Indeed, for a given $\rm Lu$, tending the magnetic field to zero along
a ray $\omega_A=\omega_{\eta}{\rm Lu}$ in the $(\omega_A,\omega_{\eta})$-plane
results in a value of the Rossby number specified by Eq. \rf{e6},
see Fig.~\ref{fig1}(c). The limit value of the critical Rossby number
oscillates between the ideal MHD value ${\rm Ro}_{\rm c}=0$
for ${\rm Lu}=\infty$ $(\varphi=0)$
and the non-magnetic value
${\rm Ro}_{\rm c}=-1-(2{\rm Re})^{-2}$ for
${\rm Lu}=0$ $(\varphi=\pi/2)$,
which explains the Velikhov-Chandrasekhar paradox.

Now we turn over to the paradox of inductionless HMRI which is
related to a similar geometric singularity as discussed above. The
leading
order WKB equations that describe the onset of instability of a
hydrodynamically stable
TC-flow with a helical external magnetic field are $\dot \xi = H \xi$ with $\xi^T=(u_R,u_{\phi},B_R(\mu_0\rho)^{-1/2},B_{\phi}(\mu_0\rho)^{-1/2})$ and
\be{e7}
H{=}\left(
      \begin{array}{cccc}
        -\omega_{\nu} & 2\Omega_0 \alpha^2 & {i\omega_{A}} & -2\omega_{A_{\phi}}{\alpha^2} \\
        -2\Omega_0(1{+}{\rm Ro})  & -\omega_{\nu} & 0 & {i\omega_{A}} \\
        {i\omega_{A}} & 0 & -\omega_{\eta} & 0 \\
        2\omega_{A_{\phi}} & {i\omega_{A}} & 2\Omega_0{\rm Ro} & -\omega_{\eta} \\
      \end{array}
    \right),
\ee
where the additional parameter
$\omega_{A_{\phi}}=R_0^{-1}B_{\phi}^0(\mu_0\rho)^{-1/2}$ is the
Alfv\'en frequency of the azimuthal magnetic field component
\cite{ks10}.
For $\omega_{A_{\phi}}=0$ these equations yield \rf{e1}.

The dispersion equation $\det(H-\gamma I)=0$  reads
\be{e8}
\lambda^4+a_1\lambda^3+a_2\lambda^2+(a_3+ib_3)\lambda+a_4+ib_4=0,
\ee
where $I$ is a $4 \times 4$ unit matrix, $\lambda=\gamma(\omega_{\nu}\omega_{\eta})^{-1/2}$, and
\ba{e8a}
a_1&=&2(1 +{\rm Pm}^{-1})\sqrt{\rm Pm},\\
a_2&=&2({1+(1+2{\beta}^2){\rm Ha}}^2)+4{\rm Re}^2   (1+{\rm Ro}){\rm Pm}+a_1^2/4, \nn \\
a_3&=&a_1(1+(1+2{\beta}^2){{\rm Ha}}^2)+8{{\rm Re}}^2(1+{\rm Ro})\sqrt{\rm Pm},\nn \\
a_4
&=&\left(1{+}{{\rm Ha}}^2\right)^2{+}4{\beta}^2{{\rm Ha}}^2
{+}4{\rm Re}^2(1{+}{\rm Ro}({\rm Pm}{{\rm Ha}}^2+1)),\nn \\
b_3&=&-8\beta{{\rm Ha}}^2{\rm Re}\sqrt{\rm Pm},~~
b_4{=}b_3(1{+}(1{-}{\rm Pm}){\rm Ro}/2)/\sqrt{\rm Pm},\nn
\ea
 where we have introduced now the Hartmann number
 ${\rm Ha}={\rm Lu}{\rm Pm}^{-1/2}$ and the helicity parameter $\beta=\alpha \omega_{A_{\phi}}\omega_A^{-1}$
of the external magnetic field. According to the analogue of the Routh-Hurwitz
conditions for the complex polynomials---the Bilharz criterion \cite{Bi44}---
the threshold of HMRI is defined by
$m_4(\beta,{\rm Re},{\rm Ha},{\rm Pm},{\rm Ro})=0$, where
$m_4$ is the determinant of the so-called
Bilharz matrix \cite{Bi44,ks10,sm11} composed of the
coefficients \rf{e8a}.
The stability condition $\Re \lambda <0$ holds if and only if $m_4>0$ \cite{Bi44,ks10,sm11}. For $\beta=0$ the dispersion equation and thus the threshold for HMRI reduce to that of
SMRI \cite{ks10}.

\begin{figure}
    \begin{center}
    \includegraphics[angle=0, width=0.3\textwidth]{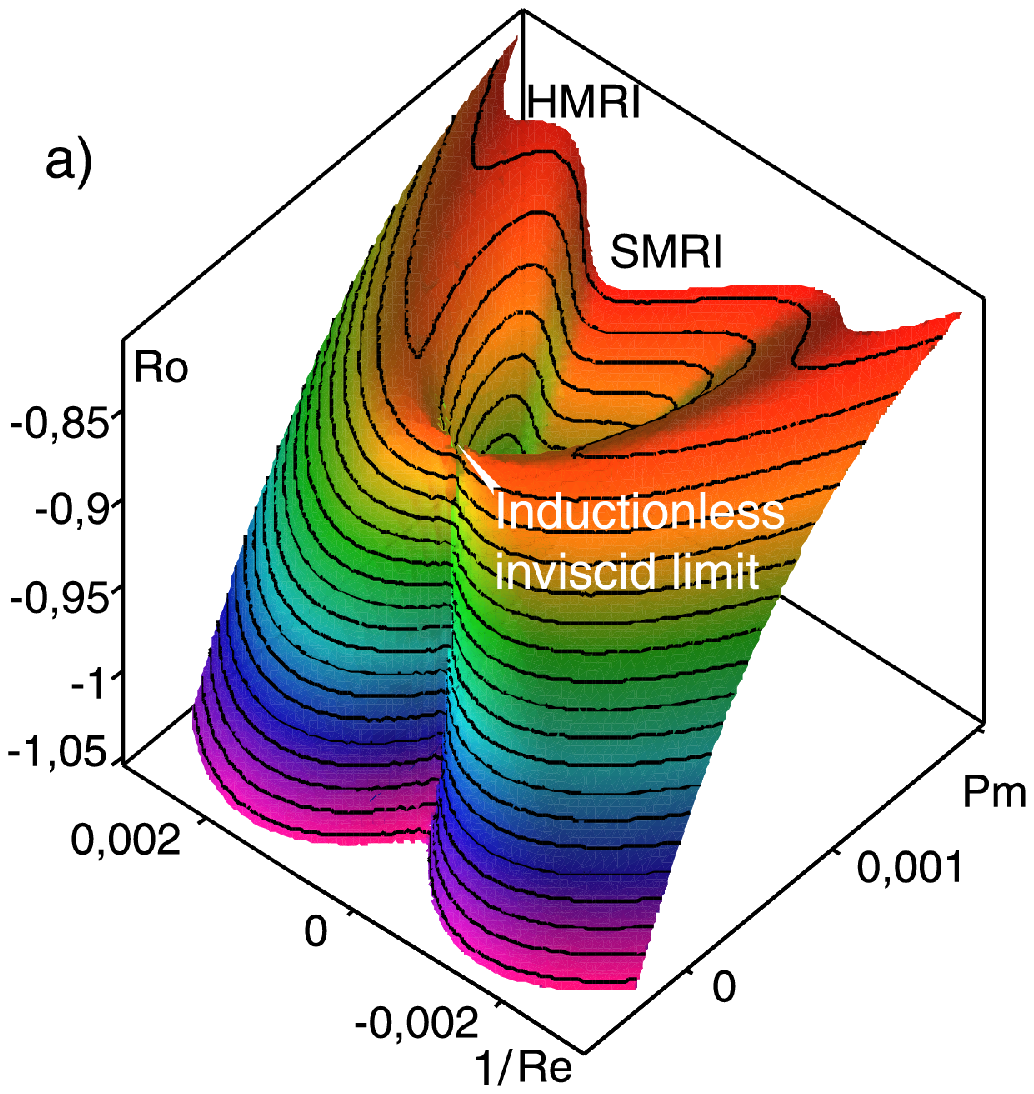}
    \hspace{.03in}
    \includegraphics[angle=0, width=0.16\textwidth]{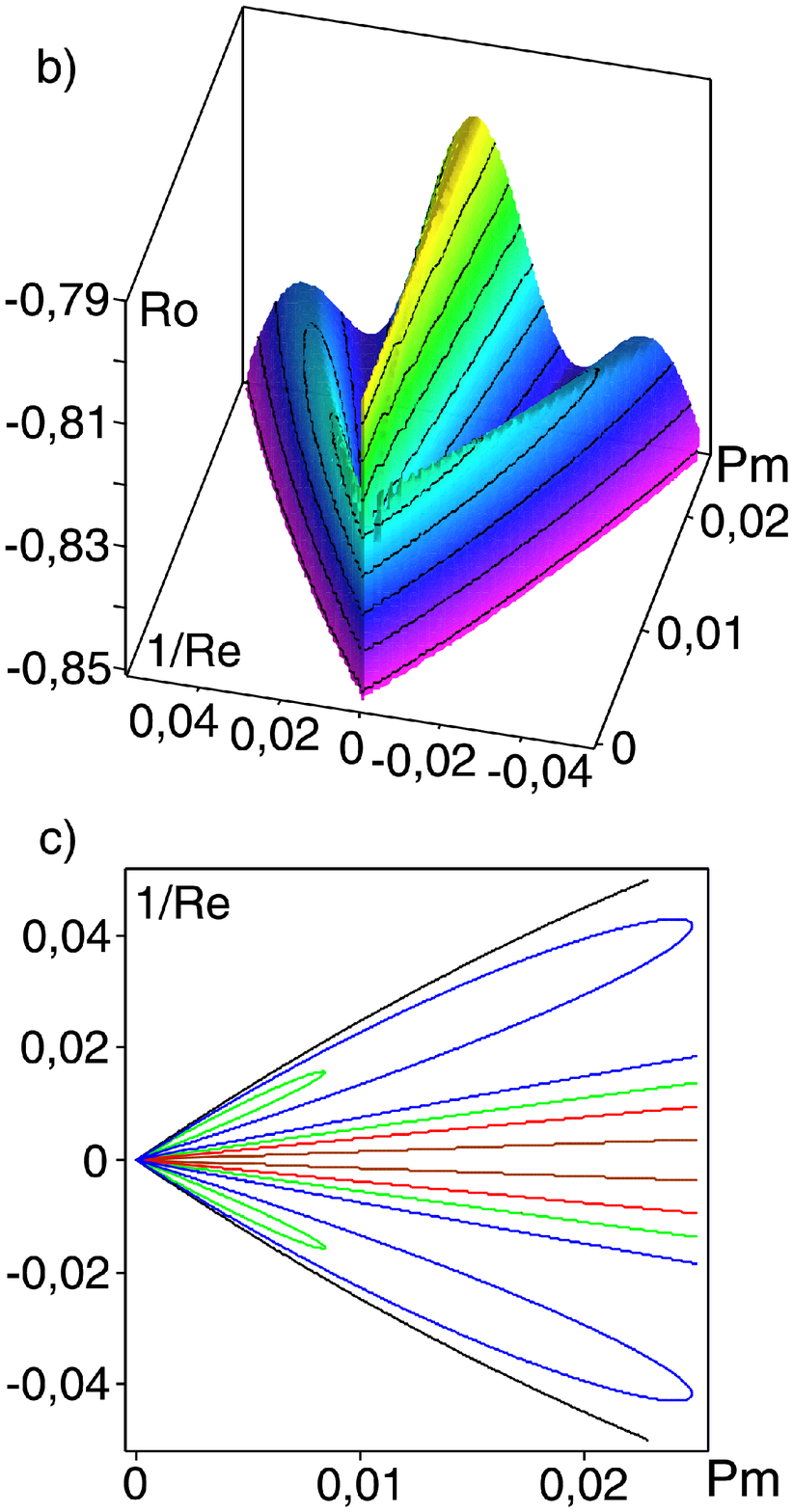}
    \end{center}
    \caption{(a) The critical Rossby number of the essential HMRI and helically modified SMRI  for ${{\rm Ha}}=15$ and $\beta=0.7$ (left) in the $({\rm Pm},{{\rm Re}}^{-1},{\rm Ro})$-space.
    (b) The critical Rossby number for ${{\rm Lu}}=0.5$ and $\beta=0.6$
    in the $({\rm Pm},{{\rm Re}}^{-1},{\rm Ro})$-space and (c) its cross-sections in the
$({\rm Pm},{{\rm Re}}^{-1})$-plane for (black) {\rm Ro}=-0.842, (blue) {\rm Ro}=
-0.832, (green) {\rm Ro}=-0.822, (red) {\rm Ro}=-0.812, (brown) {\rm Ro}=-0.802.}
    \label{fig2}
    \end{figure}

In the following we will see that in the limit
${\rm Pm} \rightarrow 0$ it is again  ${\rm Lu}$ that governs the value of ${\rm Ro}_{\rm c}$.
For this purpose, we show  in Fig.~\ref{fig2}(a)
a typical critical
surface $m_4=0$ in the $({\rm Pm},{{
\rm Re}}^{-1},{\rm Ro})$-space for the special
parameter choice ${\rm Ha}=15$ and $\beta=0.7$.
On the $\rm Ro$-axis we find a self-intersection and
two Whitney umbrella
singularities at its ends. At the upper singular point, i.e.
exactly at ${\rm Pm}=0$, we get (see \cite{ks10})
\ba{r1}
& &{\rm Ro}_{\rm c}(\beta,{\rm Ha})= \frac{\left(1{+}{\rm Ha}^2\right)^2{+}4{\beta}^2{\rm Ha}^2(1{+}{\beta}^2 {\rm Ha}^2)}{2{\beta}^2{\rm Ha}^4}\\
&-&\frac{((2{\beta}^2 {+}1){\rm Ha}^2{+}1)
\sqrt{\left(1{+}{\rm Ha}^2\right)^2{+}4{\beta}^2{\rm Ha}^2(1{+}{\beta}^2{\rm Ha}^2)
}}{2{\rm Ha}^4{\beta}^2}.\nn
\ea
In the limit ${\rm Ha}\rightarrow\infty$, this critical
value is majorated by
\be{r2}
{\rm Ro}_{\rm c}(\beta)=
\frac{1+4{\beta}^4-(1+2{\beta}^2)\sqrt{1+4{\beta}^4}}{2{\beta}^2},
\ee
with the maximum at the well-known Liu
limit ${\rm Ro}_{\rm c}=2-2\sqrt{2}\simeq-0.828$ when $\beta=\sqrt{2}/2\simeq0.707$ \cite{lghj06,ks10}.

In Fig.~\ref{fig2}(a) we see that the case with ${\rm Pm}=0$ is
connected to the case ${\rm Pm} \ne 0$
by the Pl\"ucker conoid singularity, quite similar as it was
discussed
for the paradox of Velikhov and Chandrasekhar.
Interestingly, ${\rm Ro}_{\rm c}$ for the onset of
HMRI can
increase when ${\rm Pm}$ departs from zero which happens
along curved pockets of HMRI, see Fig.~\ref{fig2}(a).
The two side bumps of the curve ${\rm Re}^{-1}({\rm Pm})$ in a
horizontal slice of the surface correspond to the domains
of the {\it essential HMRI} while the central hill
marks the {\it helically modified SMRI} domain, according to
the classification introduced in \cite{ks10}.
For small $\rm Pm$ the essential HMRI occurs at higher
${\rm Ro}$ than the helically modified SMRI, while for
some finite value of ${\rm Pm}$
the central hill and the side bumps
get the same value of $\rm Ro_c$.
Most remarkably, there is a value of $\rm Ro_c$
at which the two side bumps of
the curve ${\rm Re}^{-1}({\rm Pm})$ disappear completely.
This is the
maximal possible value for the essential HMRI,
at least at the
given $\beta$ and ${\rm Ha}$. Now we can ask: how does this
limit behave if we send ${\rm Ha}$ to infinity, and to which value
of ${\rm Lu}$ does this correspond?

Actually, with the increase in $\rm Ha$ the stability boundary
preserves its shape and simultaneously it
compresses in the direction of zero $\rm Pm$.
Substituting ${\rm Ha}={\rm Lu} {\rm Pm}^{-1/2}$ into the
equations \rf{e8a}, we plot again the surface
$m_4=0$ in the $({\rm Pm},{{\rm Re}}^{-1},{\rm Ro})$-space, but now
for a given $\beta$ and $\rm Lu$, Fig.~\ref{fig2}(b).

The corresponding cross-sections of the instability
domain in the $({\rm Re}^{-1},{\rm Pm})$-plane are shown in
Fig.~\ref{fig2}(c).
At a given value of ${\rm Ro}$ there exist three domains
of instability with the boundaries shown in blue and
green. Two sub-domains that have a form of a petal correspond to the
HMRI. They are
bounded by closed curves with a
self-intersection singularity
at the origin.
They are also elongated in a preferred direction that in
the $({\rm Re}^{-1},{\rm Pm})$-plane corresponds to a
limited range of the magnetic Reynolds number
${\rm Rm}={\rm Pm}{\rm Re}$.
The central domain, which corresponds to the
helically modified SMRI, has
a similar singularity at the origin and is unbounded
in the positive ${\rm Pm}$-direction.
In comparison with the central domain, the side petals
have lower values of  ${\rm Rm}$.

Now we reconsider again the limit ${\rm Pm} \rightarrow 0$, while keeping
$\rm Lu$ as a free parameter.
At the origin all the boundaries of the petals can be
approximated by the straight lines
$
{\rm Pm}={\rm Rm}{{\rm Re}}^{-1}.
$
Substituting this expression into equation $m_4=0$, we find that
the only term that does not depend on ${\rm Pm}$ is a polynomial
$
Q({\rm Rm},{\rm Lu},\beta,{\rm Ro})=p_0+p_1{\rm Rm}^2+p_2 {\rm Rm}^4+p_3{\rm Rm}^6,
$
where, e.g.,
$p_0={{\rm Lu}}^4(4{\beta}^4{{\rm Lu}}^2+2{\beta}^2+4{{\rm Lu}}^2{\beta}^2+1)^2$ \cite{sm11}.

\begin{figure}
    \begin{center}
    \includegraphics[angle=0, width=0.49\textwidth]{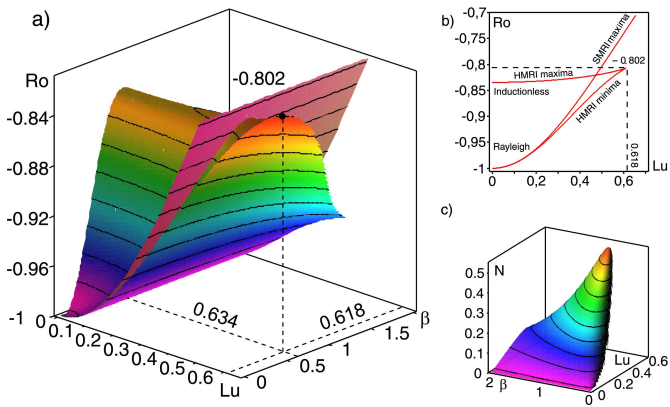}
    \end{center}
    \caption{ (a) Discriminant surface in the $({\rm Lu},\beta,{\rm Ro})$-space and (b) its cross-section at $\beta=0.634$.
    (c) Interaction parameter ${\rm N}={\rm Lu}^2 {\rm Rm}^{-1}$ at the essential HMRI maxima.}
    \label{fig3}
    \end{figure}

The roots of the polynomial are coefficients ${\rm Rm}$ of the linear
approximation to the instability domains at the origin in the
$({\rm Re}^{-1},{\rm Pm})$-plane. Simple roots mean
non-degenerate self-intersection of the stability boundary
at the origin. Double roots correspond to a degeneration
of the angle of the self-intersection when it collapses to
zero which happens only at the maximal critical Rossby number,
Fig.~\ref{fig2}(b). In the $({\rm Lu}, \beta, {\rm Ro})$-space
a set of points that correspond to multiple roots of the
polynomial $Q$ is given by the discriminant surface $64\Delta^2p_0p_3=0$ \cite{sm11}.
The surface $p_3=0$ consists of a sheet ${\rm Ro}=-(1+{\rm Lu}^2)^{-1}$
corresponding to the doubly degenerate infinite values of ${\rm Rm}$
at the maxima of the helically modified SMRI. It smoothly touches along
the $\beta$-axis the surface $\Delta=0$ that consists of two smooth sheets
that touch each other along a spatial curve --- the cuspidal edge ---
corresponding to triple roots of the polynomial $Q$,
Fig.~\ref{fig3}(a).

Every point on the upper sheet of the surface $\Delta=0$
represents a degenerate linear approximation to the essential HMRI
domain and therefore a maximal $\rm Ro$ at the corresponding values of
$\beta$ and ${\rm Lu}$. Numerical optimization
results in the new ultimate limit for HMRI
${\rm Ro}_{\rm c} \simeq-0.802$ at ${{\rm Lu}}\simeq0.618$,
$\beta\simeq0.634$, and ${\rm Rm}\simeq0.770$, see Fig.~\ref{fig3}(b).
This new limit of ${\rm Ro}_{\rm c}$ on the cuspidal edge is smoothly connected to the
inductionless Liu limit by
the upper sheet of the discriminant surface, which converges to the curve \rf{r2}
when ${\rm Lu}=0$.
We point out that the new limit is achieved at
${\rm Ha}\rightarrow \infty$ when the
optimal $\rm Pm$ tends to zero in such a way that
${{\rm Lu}}\simeq0.618$.
Figure ~\ref{fig3}(c) shows the behaviour of the so-called
interaction parameter (or Elsasser number) $\rm N=Lu^2/Rm$ for the
HMRI sheet. It is remarkable that, at $\rm Lu=0$, HMRI starts to work
already at $\rm N=0$. This can be explained by the observation that
the optimal value for HMRI corresponds to
${\rm N}{\rm Ha}={\rm Lu}^3 /({\rm Rm}\sqrt{\rm Pm})=1/(1+2^{-1/2})=0.586$,
\cite{ks10}. Later, for increasing
$\rm Lu$, the optimal ${\rm N}$ acquires final values, passes through its maximum
and at ${{\rm Lu}}\simeq0.618$ and
$\beta\simeq0.634$ it terminates at $\rm N=0.496$.

Inspired by the theory of dissipation induced instabilities
\cite{kv10}, we
have resolved the two paradoxes of SMRI and HMRI in the
limits of infinite and zero magnetic Prandtl number, respectively,
by establishing their sharp correspondence to singularities on the instability thresholds.
In either case, it is the local Pl\"ucker conoid structure that
explains the non-uniqueness of the critical Rossby number, and its
crucial dependence on the Lundquist number. For HMRI, we have found an extension
of the former Liu limit ${\rm Ro}_{\rm c} \simeq -0.828$
(valid for ${\rm Lu}=0$) to a somewhat higher
value ${\rm Ro}\simeq-0.802$ at ${\rm Lu}=0.618$ which is, however,
still below the Kepler value.
To study the possible consequences of this new limit for the
saturation of MRI in accretion disks or experiments, is left for future work.

\begin{acknowledgments}
Financial support from the Alexander von Humboldt Foundation and the DFG
in frame of STE 991/1-1 and of SFB 609 is gratefully acknowledged.
\end{acknowledgments}

\end{document}